\documentclass{ws-ijbc}
\usepackage{ws-rotating}     
\usepackage{graphicx}
\usepackage{epstopdf}
\usepackage{booktabs}
\newcommand{\ra}[1]{\renewcommand{\arraystretch}{#1}}

\ifx\pdfoutput\@undefined\usepackage[usenames,dvips]{xcolor}
\else\usepackage[usenames,dvipsnames]{xcolor}
\IfFileExists{pdfcolmk.sty}{\usepackage{pdfcolmk}}{} 
\fi
\usepackage[plainpages=false,pdfpagelabels,pagebackref=false,naturalnames=true,hyperindex=true,pdftitle={Asymptotic Behaviour and Ratios of Complexity in Cellular Automata},pdfauthor={Hector Zenil}]{hyperref}
\hypersetup{colorlinks=true,
urlcolor=Cerulean,linkcolor=BrickRed,citecolor=RoyalBlue,
  pdfpagemode=None,
  pdfstartview=FitH}
\usepackage[all]{hypcap}

\begin{document}

\catchline{}{}{}{}{} 

\markboth{H. Zenil}{Asymptotic Behaviour and Ratios of Complexity in Cellular Automata}

\title{ASYMPTOTIC BEHAVIOUR AND RATIOS OF COMPLEXITY IN CELLULAR AUTOMATA}

\author{HECTOR ZENIL}

\address{Department of Computer Science / Kroto Research Institute, University of Sheffield, Sheffield, S. Yorkshire, S1 4DP, United Kingdom.\\
hectorz@labores.eu}

\maketitle


\begin{abstract}
We study the asymptotic behaviour of symbolic computing systems, notably one-dimensional cellular automata (CA), in order to ascertain whether and at what rate the number of complex versus simple rules dominate the rule space for increasing neighbourhood range and number of symbols (or colours), and how different behaviour is distributed in the spaces of different cellular automata formalisms. Using two different measures, Shannon's block entropy and Kolmogorov complexity, the latter approximated by two different methods (lossless compressibility and block decomposition), we arrive at the same trend of larger complex behavioural fractions. We also advance a notion of asymptotic and limit behaviour for individual rules, both over initial conditions and runtimes, and we provide a formalisation of Wolfram's classification as a limit function in terms of Kolmogorov complexity.
\end{abstract}


\section{Introduction}

We address several open questions formulated in~\cite{wolframopen,wolfram1985} concerning the distribution of different behaviours in the space of cellular automata rules when rule spaces increase in size by number of states and symbols. The definite answer is ultimately uncomputable (see~\cite{kn:CY88}), as a behavioural characterisation of all its future states requires a full and detailed understanding of the dynamics of the system, something that is ultimately impossible due to the halting and reachability problems. An introduction to open questions in computability and cellular automata can be found in~\cite{acomputableuniverse}. Here we provide a numerical approach to both the question of behavioural changes of a system over time and the question of the fraction of complex versus simple behaviour in a rule space, using 2 measures of complexity derived from algorithmic information theory and proven to have the power to characterise randomness. We provide statistics on fractions of complex behaviour in cellular automata when the number of symbols and neighbourhood ranges increase.

\section{Cellular Automata}

Let $S$ be a finite set of symbols (or colours) of a cellular automaton (CA). A finite configuration is a configuration with a finite number of symbols which differ from a distinguished state $b$ (the grid \emph{background}) denoted by $0^\infty b 0^\infty$ where $b$ is a sequence of symbols in $S$ (if binary then $S=\{0,1\}$). A stack of configurations in which each configuration is obtained from the preceding one by applying the updating rule is called an evolution. Formally:

Let $f:S^\mathbb{Z} \rightarrow S^\mathbb{Z}$ and $n,i \in \mathbb{N}$ then:

\begin{equation}
f(r_t) = \lambda(x_{i-r} \ldots x_i \ldots x_{i+r})
\end{equation}

Where $f$ is a configuration of the CA and $r_t$ a row with $t\in\mathbb{N}$ and $r_0$ the initial configuration (or initial condition). $f$ is also called the global rule of the CA, with $\lambda:S^n \rightarrow S$ the local rule determining the values of each cell and $r$ the neighbourhood range or radius of the cellular automaton, that is, the number of cells taken into consideration to the left and right of a central cell $x_i$ in the rule that determines the value of the next cell $x^\prime_i$. All cells update their states synchronously. Cells at the extreme end of a row must be connected to cells at the extreme right of a row in order for $f$ to be considered well defined.

\subsection{Elementary Cellular Automata}

The simplest non-trivial CA rule space is the nearest neighbourhood ($r=1$) one-dimensional CA with two possible symbols (or colours) per cell. These are called Elementary Cellular Automata (ECA) as defined in~\cite{wolfram}, where $\lambda:S^3 \rightarrow S$. As for general CA, the set of local rules specifies the updated value of a site for each possible configuration. In an ECA, each site takes one of two values, conventionally denoted by ``0'' and ``1'' (graphically depicted as white and black cells, respectively). There are exactly $2^{2^{2r+1}}=256$ rules of this type.

\subsection{Totalistic Cellular Automata}

Given that the number of rules for general cellular automata grow too fast for an increase in the number of $k$ symbols (colours) and range $r$ (neighbourhood) values, we also decided to study the behavioural fractions of totalistic CAs (denoted by $CA_T$) in complementation to the limited study of general CA (starting from ECA) as a rule space to investigate, given that because of their rule-averaged nature their rule spaces grow more slowly than those of general CA rules. This is because the equation $x_{i-m} + x_i + x_{i+n}$ can represent the same value despite different combinations of $x_i$.

A totalistic cellular automaton is a cellular automata in which the rules depend only on the total (or equivalently, the average) of the values of the cells in a neighbourhood. Formally, the global rule of a totalistic CA can be defined by:

\begin{equation}
f(x_i) = \lambda(x_{i-m} + x_i + x_{i+n})
\end{equation}

\noindent Where as before, $x_i$ is a cell in the CA grid and $\lambda$ the local rule. A fractional ratio such as $r=2/3$ means that the rule takes $m=2$ cells to the left (including the central one) and $n=3$ to the right (including the central one). A full introduction to these automata can be found in~\cite{wolfram}. Like an ECA, the evolution of a one-dimensional totalistic cellular automaton ($CA_T$) can be fully described by specifying the state a given cell will assume in the next generation based on the average value of the cells in the neighbourhood, consisting of the cells to its left, the value of the cell itself, and the values of the cells to the right, according to the neighbourhood range $r$. The best known, albeit two-dimensional, totalistic cellular automaton is Conway's cellular automaton known as the Game of Life~\cite{life}. The total number of rules for totalistic CA is given by:

\begin{equation}
CA_T(k,r)=k^{((k - 1) (2 r + 1) + 1)}
\end{equation}

\noindent With $k$ the number of symbols (colours) and $r$ the range or neighbourhood.

\section{Wolfram's Classification}
\label{wolframclassification}

Analysing the average performance of a program is a key problem in computer science. Wolfram advanced~\cite{wolfram} an heuristic for classifying computer programs by their space-time diagrams. Computer programs behave differently for different inputs. It is possible and not uncommon, however, to analyse a program with respect to a single fixed input, for example, based on a counting argument from algorithmic information theory. The counting argument is based on the Pigeonhole principle and tells us that most strings are Kolmogorov random~\cite{li}, simply because there are significantly fewer shorter binary programs than binary strings of a fixed size. The approach is very similar to the Incompressibility Method~\cite{li} and may justify the study of the evolution of a computer program from a random single input as an average case. However, here we are also interested in the rates of change of behaviour of single cases and in the comparative rates of change of all computer programs in a defined space.

\begin{figure}[h!]
\begin{center}
\scalebox{.4}{\includegraphics{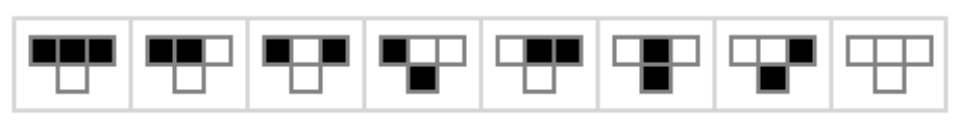}}
\scalebox{.26}{\includegraphics{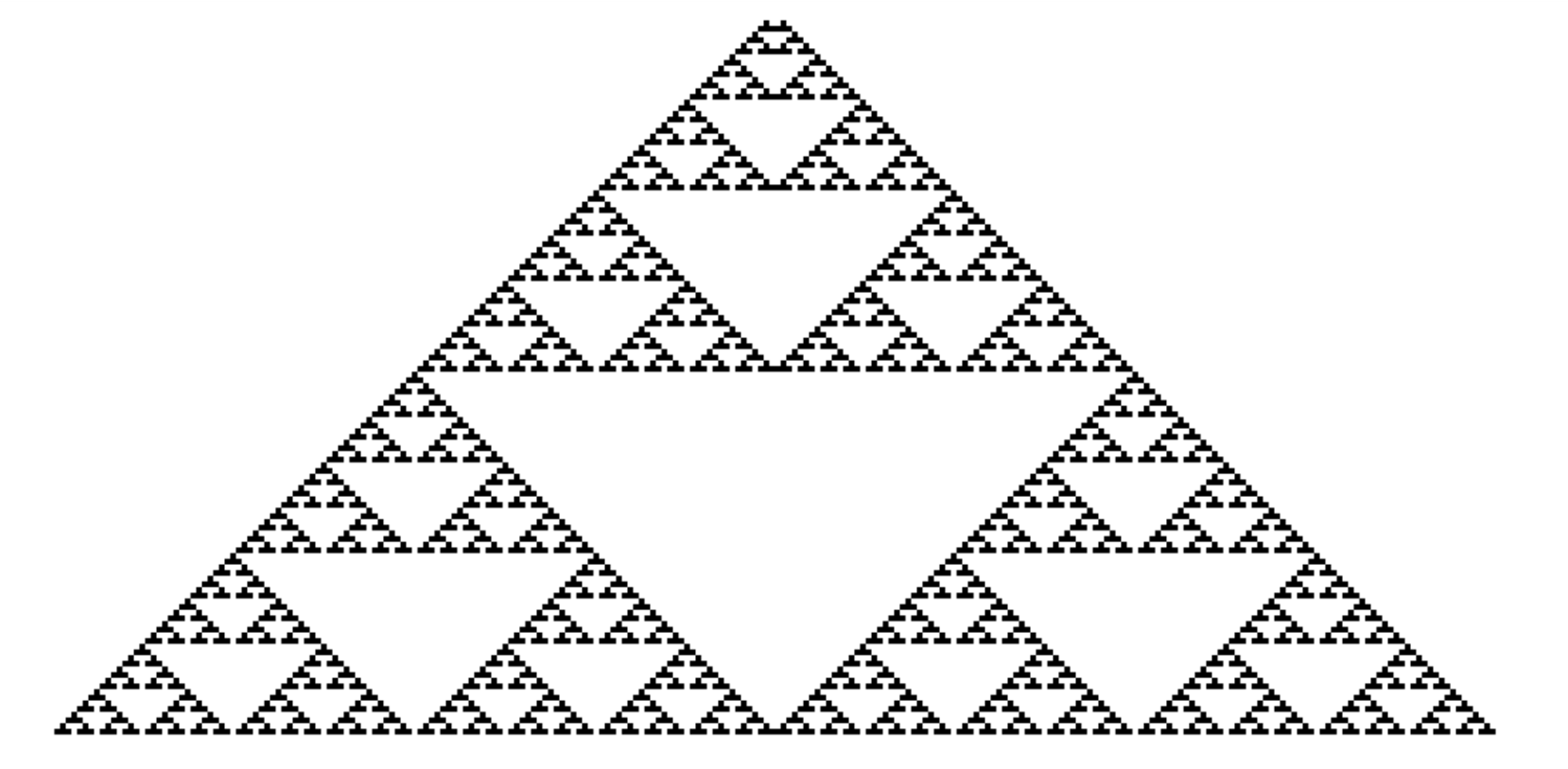}}\scalebox{.26}{\includegraphics{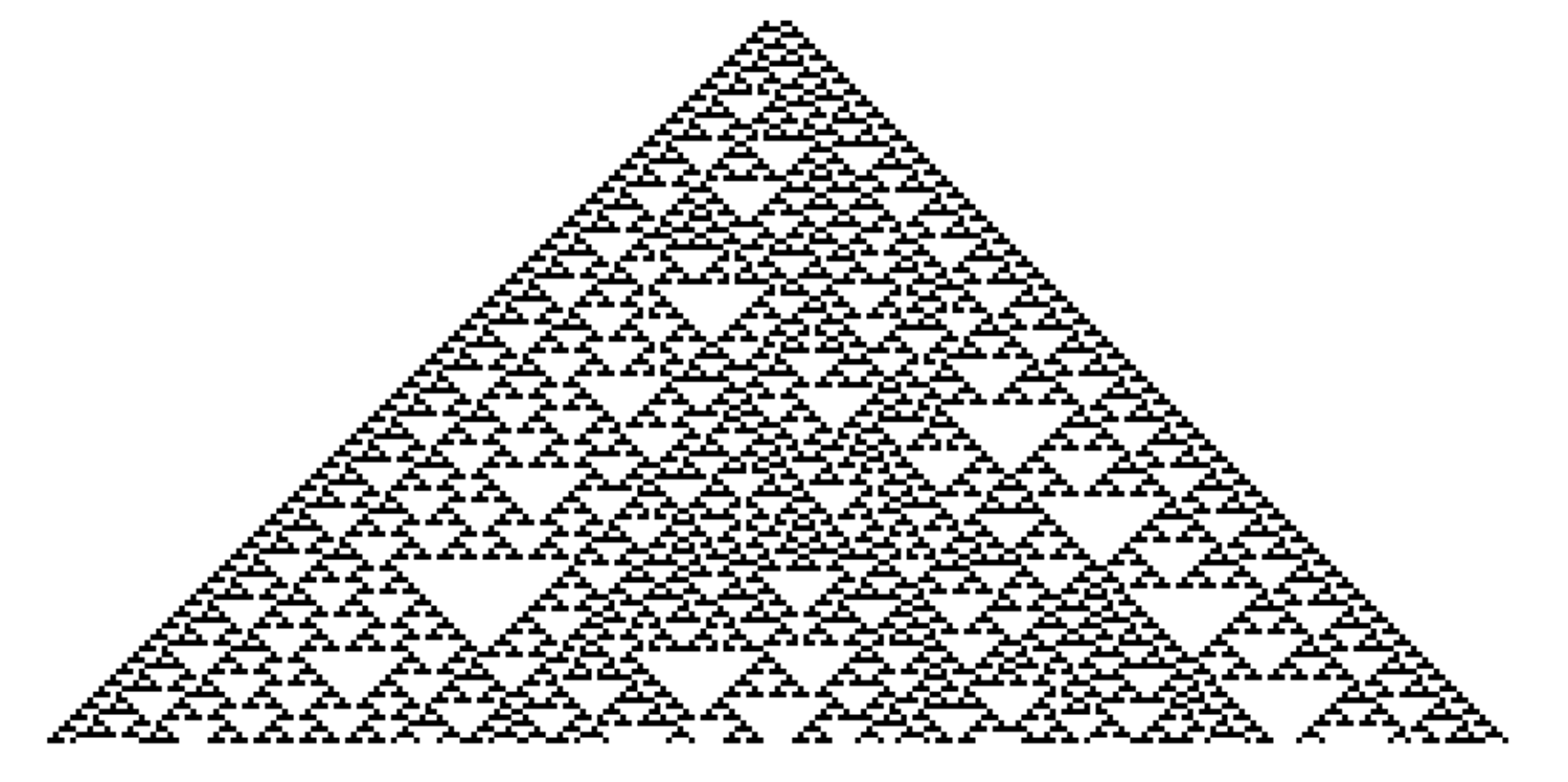}}
\end{center}
\caption{Rule 22 starting from initial configuration 1001 (Left) versus starting from 1011 (Right).}
\label{rule22a}
\end{figure}

Evolutions of computer programs can display a variety of different qualitative properties (see, for example, Fig.~\ref{rule22a}), some of which display behaviour associated with pseudo-randomness even for simple initial configurations~\cite{wolfram}. Others have also been proved to be capable of Turing universal computation~\cite{cook,wolfram}.

An heuristic method to classify behavior of computer programs was introduced by Wolfram~\cite{wolfram} by examining their space-time evolutions starting with a random initial configuration. Wolfram's classes can be characterised as follows:

\begin{itemlist}
\item {Class 1.} Symbolic systems which rapidly converge to a uniform state. Examples are rules 0, 32 and 160.
\item {Class 2.} Symbolic systems which rapidly converge to a repetitive or stable state. Examples are rules 4, 108 and 218.
\item {Class 3.} Symbolic systems which appear to remain in a random state. Examples are rules 22, 30, 126 and 189.
\item {Class 4.} Symbolic systems which form areas of repetitive or stable states, but which also form structures that interact with each other in complicated ways. Examples are rules 54 and 110.
\end{itemlist}

Random sampling yields some empirical indications of the frequencies of different classes of behaviour among cellular automaton rules of various kinds. The choice of a ``random'' initial configuration is a way to capture the ``average'' behaviour of a system, given that if a CA behaves in some determined fashion for most initial configurations, there is a better chance that a ``random'' initial configuration will characterise the typical behaviour of said CA. 

There has been a significant amount of work done on the classification of cellular automata in e.g.~\cite{langton}, ~\cite{gutowitz}, ~\cite{wuensche}, among others. A good survey and introduction can be found in~\cite{genaroca}. Wuensche, for example, looked for signatures to aid searches, and statistics on changing portions in each class as colours and neighbourhood ranges increased. The problem here is that a particular initial condition may yield atypical results, so the results need to be averaged over a sample of initial conditions. But an optimal definition would capture whether the system actually reaches a quintessential state showing a stable evolution over time. Another variable is of course the number of time-steps before measures start to allow the system to settle into typical behaviour, as was also explored in~\cite{wuensche}. Here, however, we provide evidence that systems tend to settle into their typical behaviours very soon.

Let us define a recursive function $W$ that retrieves the Wolfram class of a computer program $s$ represented by its space-time evolution (hence effectively a string---e.g. for a Turing machine---or an $n$-dimensional array---e.g. for a cellular automaton). A Wolfram class $W$ of a system $s$ is recursively given by

$$W(s)=\lim_{i,t\rightarrow\infty}\max{W(s(i,t))}$$

\noindent where $W(s(i,t))$ is the complexity Wolfram class of $s$ for input with index $i$ (for an arbitrary enumeration) up to runtime $t$ and $W(s)$ is the maximum complexity class over the set $\{W(s(i,t))\}$ for all $i$ and $t$. $W(s)$ is approachable from below  if $W(s(i,t))=C_n$ for some $n$ and $W(s(i^\prime,t^\prime))=C_n^\prime$ then $s$ is in Wolfram class $C_n\prime$ and not in $C_n$ if and only if $C_n\prime > C_n$. $W$  induces a non-effective partition when $t \rightarrow \infty$ over all (infinitely enumerable) computer programs $s$ where no program can belong to 2 different Wolfram classes in the limit or, more formally, $\bigcap_{n=1}^4 C_n=\emptyset$ and $\sum_{n=1}^4 |C_n| = |\bigcup_{n=1}^4 C_n|$. This also implies that one can misclassify a computer program for some values $i$ and $t$ but not in the limit.

$W(s(i,t))$ can be formalized by using a suitable complexity measure, in general this will be taken to be the maximal Kolmogorov complexity $K(s(i,t))$ for all initial conditions $i$ up to time $t$. Here, in practice this will be some lossless compression algorithm implementing Entropy rate as a test for non-randomness and therefore a loose upper bound on $K$. Thresholds are then trained to divide $W$ into 4 classes $C_n$ with $n \in \{1 ,2 ,3 ,4\}$.

We can now formally set forth the question that is of interest in this paper as follows. Because $W$ induces an equivalence class in the limit, we know that $\bigcap_{n=1}^4 C_n=\emptyset$ and $\sum_{n=1}^4 |C_n| = |\bigcup_{n=1}^4 C_n|$. What we're interested in in this paper is the contribution of each $|C_n|$ to the total sum when the available resources (e.g. states, symbols) of computer programs grow. Our particular concern is a type of program studied by Wolfram himself, that is, one-dimensional cellular automata, when the number of available symbols $k$ and neighbourhood range $n$ grow.

Given Wolfram's Principle of Computational Equivalence (PCE) and because systems in $C_{3,4}$ and $C_{1,2}$ can be said to display the most widely divergent behaviour, a simplified question we ask in this paper concerns the fraction of systems in $C_{3,4}$ with respect to $C_{1,2}$ for a finite set of computer programs (cellular automata).

\subsection{Asymptotic Behaviour of Cellular Automata}

Central to AIT is the basic definition of plain algorithmic (Kolmogorov-Chaitin or program-size) complexity~\cite{kolmo,chaitin}:

\begin{equation}
\label{kolmo}
K_U(s) = \min \{|p|, U(p)=s\}
\end{equation}

Where $U$ is a universal Turing machine and $p$ the program that, running on $U$, produces $s$. Traditionally, the way to approach the algorithmic complexity of a string has been by using lossless compression algorithms. The result of a lossless compression algorithm applied to $s$ is an upper bound of the Kolmogorov complexity of $s$.

The invariance theorem guarantees that complexity values will only diverge by a constant $c$ (e.g. the length of a compiler or a translation program) (see~\cite{calude,li}). If $U_1$ and $U_2$ are two universal Turing machines, and $K_{U_1}(s)$ and $K_{U_2}(s)$ the algorithmic complexity of $s$ for $U_1$ and $U_2$ respectively, there exists a constant $c$ such that:   $| K_{U_1}(s) - K_{U_2}(s) | < c$.

Hence the longer the string, the less important $c$ is (i.e. the choice of programming language or universal Turing machine). $K$ as a function from $s$ to $K(s)$ is upper semi-computable, meaning that one can find upper bounds. The shortest description of a string can be interpreted as a compressed version of itself. The result of a lossless compression algorithm is an upper bound of its algorithmic complexity. A compressed version of a string is actually a sufficient test of non-algorithmic randomness or low Kolmogorov complexity. 

Kolmogorov proposed to define the $c$-incompressibility of a string $s$ if $length(s)-K(s)<c$, and in~\cite{martin} that a string $s$ is $c$-random if it cannot be selected by any possible computable tests at a level of precision $c$.


\subsection{A definition of sensitivity to initial conditions}

Another question about cellular automata is the matter of their stability under small perturbations to their initial conditions. In a previous article a classification of ECAs by sensitivity to initial conditions was introduced~\cite{zenilca}. In this article we want to look at the asymptotic behaviour of the ECAs that do present phase transitions. This focus was dictated by the concept of Kolmogorov complexity. The method produces the following variation of Wolfram's classification \cite{kn:ZenAISB}:

\begin{itemlist}
\item {Classes 1 and 2 ($C_{1,2}$).} The evolution of the system is highly compressible for any number of steps;
\item {Classes 3 and 4 ($C_{3,4}$).} The lengths of the compressed evolutions asymptotically converge to the lengths of the uncompressed evolutions.
\end{itemlist}

A measure based on the change of the asymptotic direction of the size of the compressed evolutions of a system for different initial configurations (following a proposed Gray-code enumeration of initial configurations) was presented in~\cite{zenilca}. It gauges the resiliency or sensitivity of a system vis-\`a-vis its initial conditions. The phase transition coefficient defined therein led to an interesting characterisation and classification of systems, which when applied to elementary CA, yielded exactly Wolfram's four classes of system behaviour. The coefficient works by compressing the changes of the different evolutions through time, normalised by evolution space. It has proved to be an interesting way to address Wolfram's classification questions. The motivation in~\cite{zenilca} was to address one of the apparent problems of Wolfram's original classification, that of rules behaving in different ways when starting from different initial configurations. 

\begin{figure}[h!]
\begin{center}
\scalebox{.54}{\includegraphics{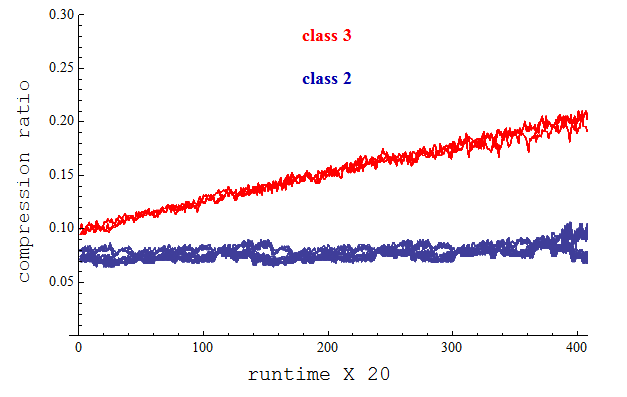}}
\end{center}
\caption{Behavioural bi-stability of Rule 22 over the first 20 initial configurations according to the Gray-code numbering scheme, running for $10^3$ steps each, and compressed every 500 steps, representing a point. The rule's behaviour is qualitatively the same according to its information-content measures by compressibility.}
\label{rule22asymp2}
\end{figure}

Rule 22, just like other ECA rules (e.g. rule 109), evolves in 2 different, clearly distinguished fashions (see Fig.~\ref{rule22a}). One is asymptotically compressible and the other asymptotically uncompressible. For example, initial configurations containing substrings of the form $1^n0^n1$ for $n>1$, where $X^n$ is a repetition of $n$ times the bit $X$, trigger a Class 3 (random looking) behaviour (see Fig.~\ref{rule22inits}), which then dominates over time, with no qualitative change the longer the system runs (see~Fig. \ref{rule22asymp2}). However, evolutions starting from symmetric initial configurations of odd length cannot lead to non-symmetric configurations after application of the local rules of ECA rule 22 hence only leading to Class 2 behaviour (periodic) leaving the system in full equilibrium. On the other hand, for evolutions of rule 22 out of equilibrium, the probability of moving from a disordered state to a symmetrical one can be calculated by the size of the evolution of the cellular automaton from which the chances to reach a symmetric configuration decays exponentially with the ECA runtime given that the number of symmetric configurations are outnumbered by the number of random strings from a simple counting argument.

\begin{figure}[h!]
\begin{center}
\scalebox{.32}{\includegraphics{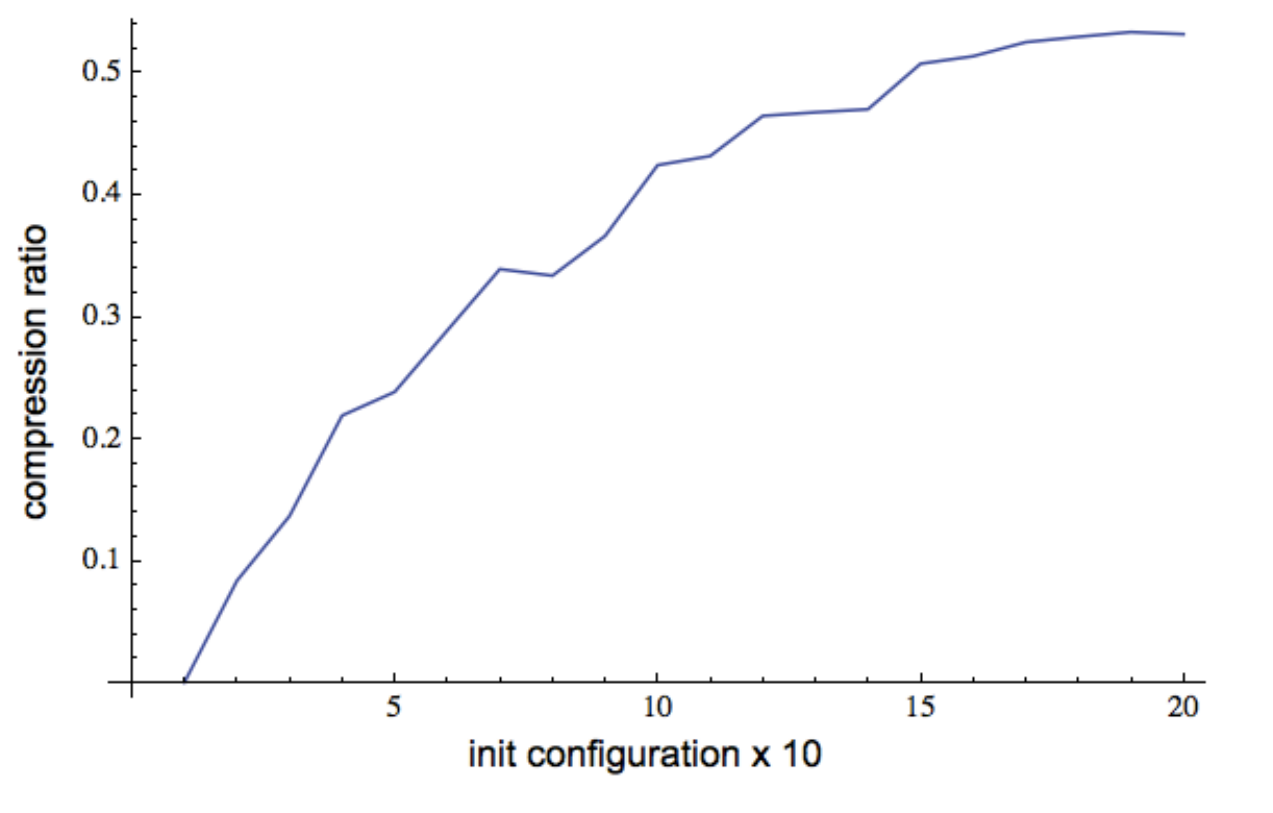}} \scalebox{.32}{\includegraphics{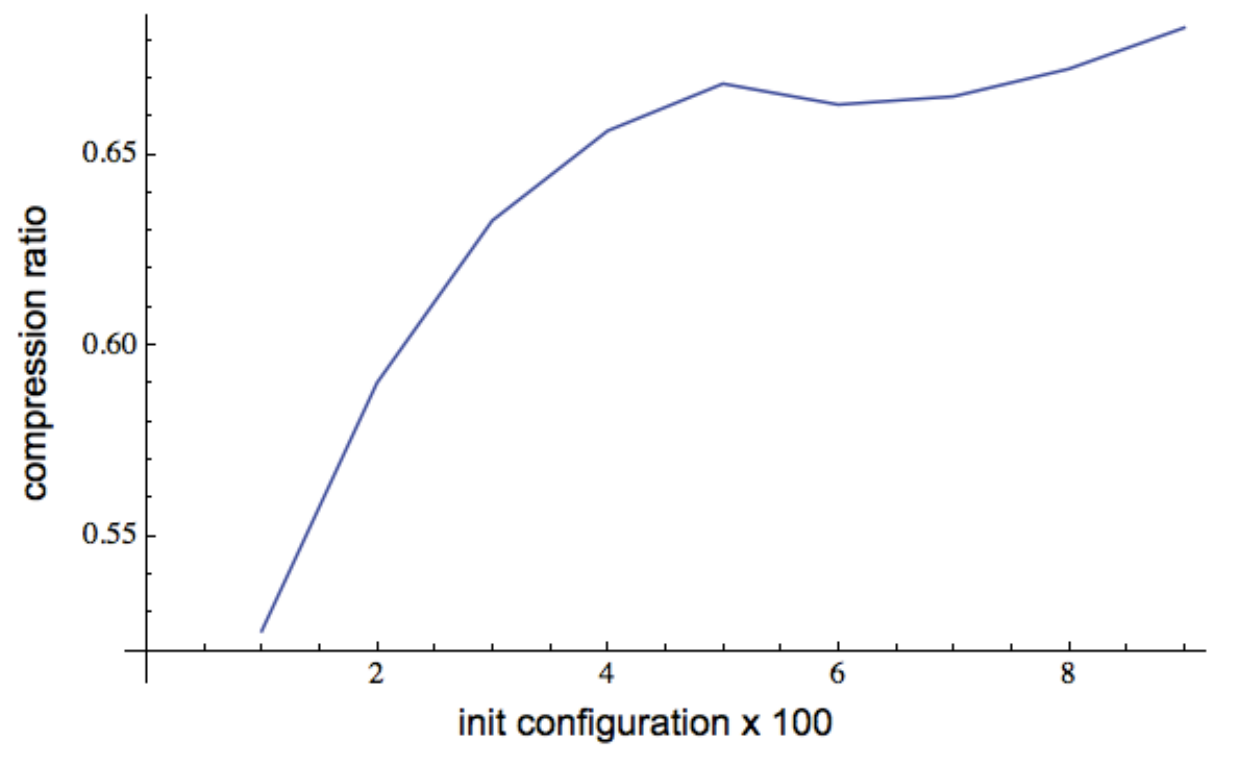}}
\end{center}
\caption{Asymptotic behaviour of rule 22. Left: First 200 initial conditions in intervals of 10. Right: Initial conditions 200 to 1000 in intervals of 100. The uncompressibility ratio grows for longer initial configurations.}
\label{rule22}
\end{figure}

Initial conditions that lead to different behaviours remain stable; no behavioural changes occur between the 2 groups of asymptotically compressible and uncompressible behaviour. By applying regression analysis over Fig.~\ref{rule22asymp2} approaching the 2 curves corresponding to the compressed versions of the behaviours of rule 22, it can be seen that the partial derivatives of the calculated fitted curves are either constant for uncompressible Class 3 behaviour or 0 for compressible Class 2 behaviour. Hence they are stable over time.

\subsubsection{Initial configuration numbering scheme and a distance metric}
\label{enumeration}

Ideally, one should feed a system with a natural sequence of initial configurations of gradually increasing complexity, so that qualitative changes in the evolution of the system are not attributable to large changes in the initial conditions.

More precisely, Gray codes minimise the Damerau-Levenshtein distance between strings. The Damerau-Levenshtein distance between two strings $u$ and $v$ gives the number of one-element deletions, insertions, substitutions or transpositions required to transform $u$ into $v$.

\begin{figure}[h!]
\begin{center}
\scalebox{.36}{\includegraphics{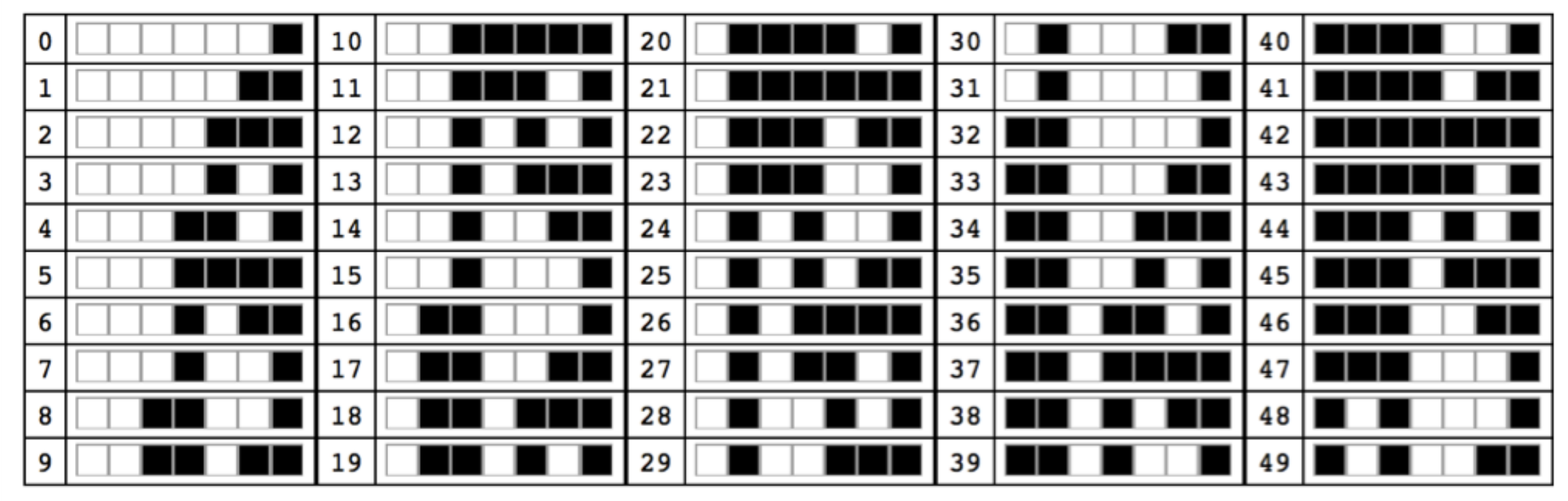}}
\end{center}
\caption{First 50 initial configurations based on a Gray-code enumeration for binary CAs.}
\label{gray}
\end{figure}

The simplest, not completely trivial, initial configuration of a cellular automaton is the typical single black cell on an ``empty'' (or white) background. An initial configuration is the region that must be varied, consisting only of the non-white portion of a system. For example, the initial configuration 010 is exactly the same as 1 because the cellular automaton background symbol is zero. Therefore valid initial configurations for cellular automata should always be wrapped in 1's as they are in Fig.~\ref{gray}.

Gray codes exist for non-binary strings. These are $n$-ary Gray codes that allow us to generate initial conditions for computing systems such as cellular automata with more than 2 states or colours.

\begin{figure}[h!]
\begin{center}
\scalebox{.5}{\includegraphics{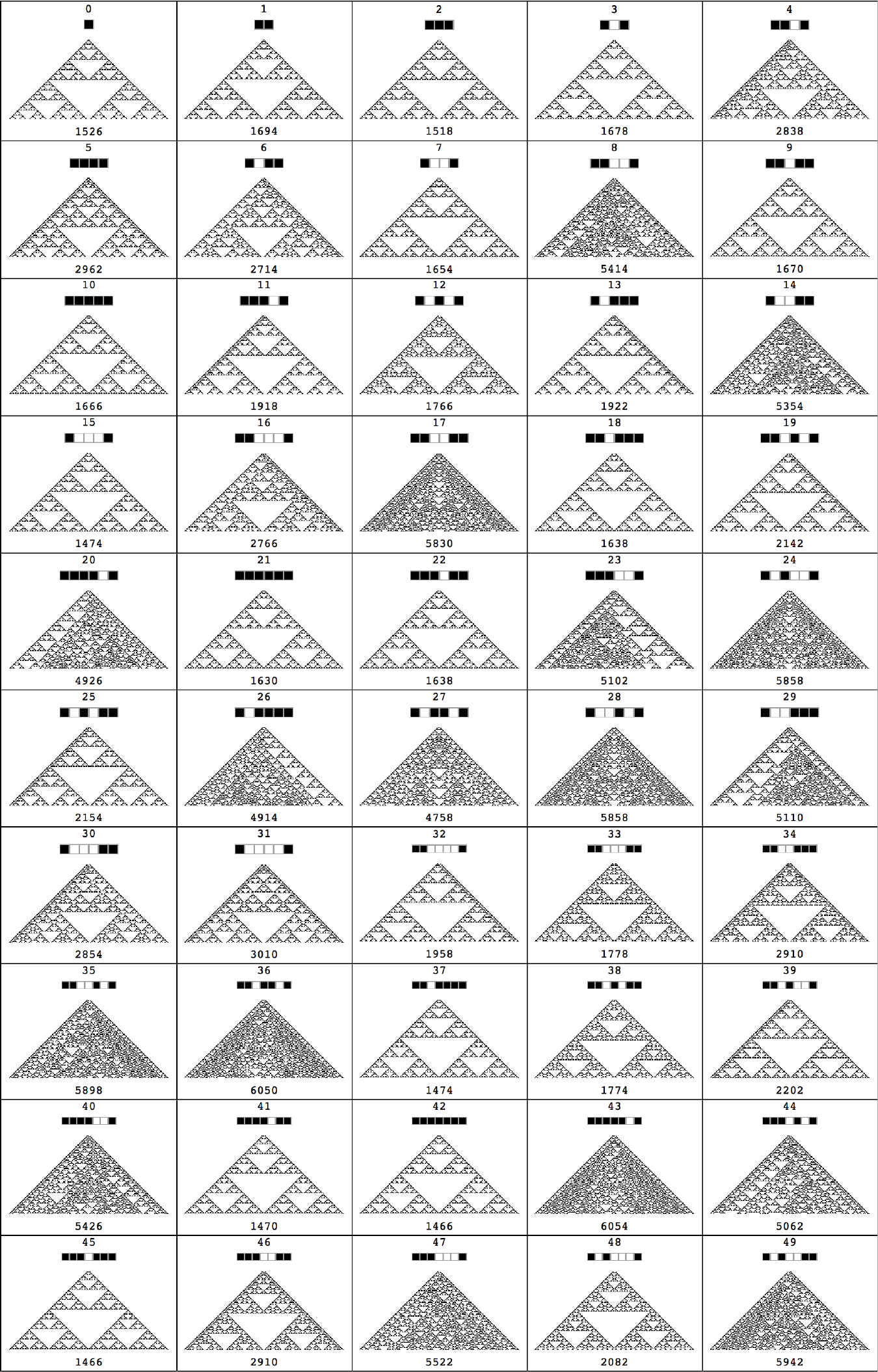}}
\end{center}
\caption{Sensitivity of Rule 22 to initial configurations under the Gray-code initial configuration scheme. Each evolution is preceded by the initial configuration and followed by the losslessly compressed (Deflate) length. Differences in patterns generated by ECA Rule 22 induced by gradual changes following the Gray-code based enumeration for CA, with each evolution running for 125 steps.}
\label{rule22inits}
\end{figure}

In Fig.~\ref{rule22inits} one observes 28 of the first 50 initial configurations leading to simple periodic evolutions, which amounts to a 56\% chance. However, many are produced from the same pattern (symmetrical inputs) and several others are mirror evolutions.  Only 16 are truly different, that is, 32\% of the first 50. Hence rule 22 is 32:50 sensitive, unlike, for example, rule 30, which, like rules such as rule 0, is highly robust, whereas rules such as rule 110 fall in the middle, being moderately sensitive. A systematic analysis of sensitive, moderate and robust systems may be found in~\cite{genaronew}.

\section{Measures of information content and algorithmic complexity}

We aim to address the question of the behaviour of a system using a definition of asymptotic and limit behaviour that is cognizant of Wolfram's classification problem (systems which may behave differently for different initial configurations), and that is unlike other approaches (e.g. see~\cite{langton} or~\cite{wuensche,wuensche2}). We follow 2 different methods. One is based on the traditional Shannon entropy, which has the advantage of being computable but lacks~\cite{grunwald} the power of a universal measure of complexity, while the other is based on Kolmogorov complexity, which has the advantage of the full power of a universal complexity measure (\cite{chaitin}, \cite{martin}, \cite{schnorr} and\cite{gacs}) but is semi-computable. To approximate Kolmogorov complexity too we used  2 different, complementary approaches, one being the traditional method using lossless compression algorithms and the other the Coding Theorem Method as defined in~\cite{d4,d5}.

\subsection{Normalised Block Shannon Entropy}

A probabilistically-based complexity measure is given by \emph{Block entropy}, which, as suggested in~\cite{wolfram}, generalizes the concept of Shannon entropy~\cite{shannon} to blocks of $N \times N$ symbols. A normalised version of the Block entropy for block size $N$ is proposed below:

\begin{equation}
\label{blockentropy}
H_N(g) = \frac{\sum_{r\in \{g\}_{N\times N}} E(r) / \log_2(N\times N)}{|g|}
\end{equation}
where $\{g\}_{N\times N}$ is the object $g$ decomposed into square blocks of length $N$ and $E(r)$ the Shannon entropy of each block $r$ (with boundary conditions). For $N = 1$, for example, block entropy is simply the standard unigram entropy. For $N = 2$, it is the entropy of 2-dimensional arrays of size $N \times N$.

This measure is used as a computable, if limited, complexity measure to provide a first indication, before proceeding to estimate Kolmogorov complexity in a bid to quantify fractions of behavioural changes in increasingly larger spaces of cellular automata.

\subsection{Normalised (lossless) Compression}

We are also interested in the following information-theoretic \emph{Normalised Compression} ($NC$) measure:

\begin{equation}
NC(s) = K(s)/length(s)
\end{equation}

\noindent Where $K(s)$ is the Kolmogorov complexity of $s$. Given that $K$ is lower semi-computable we replace $K$ by the values retrieved by a general lossless compression algorithm $C$ for which compressed lengths of $s$ are upper bounds of $K$. Large compression ratios $NC$ therefore represent a sufficient test of low Kolmogorov complexity. Because compression algorithms can retrieve values larger than the uncompressed original objects we use the following variation:

\begin{equation}
NC(s) = C(s)/\max(C(s), length(s))
\end{equation}

We distinguish complex from simple behaviour in CA by means of the lossless compressibility approximation of Kolmogorov complexity. We consider a cellular automaton to be complex or simple if the compressed lengths tend to compression ratio $NC$ 0 or 1. Formally, a system $s(i,t)$ is said to be complex for initial configuration number $i$, following a Gray code, and time $t$, if:

\begin{equation}
\lim_{i,t \rightarrow \infty} C(s(i,t)) = |s(i,t)|
\end{equation}

The question of the asymptotic behaviour of a cellular automaton $s$ is therefore the question of whether $\lim_{i \rightarrow \infty} NC(s(i)) =1$ for complex behaviour, or 0 for simple behaviour. We will use $NC$ to determine the Wolfram class of a CA $s$. If $NC(s) \approx 1$ then $W(s)=C_{3,4}$; otherwise $W(s)=C_{1,2}$. However, we need a numerical approximation of $NC(s)$, which means evaluating $NC(s(i,t)$ for a number of initial conditions $i$ following the Gray-code numbering scheme, and for a runtime $t$. Therefore the approximated Wolfram class would be $W(s(i,t))$.



An interesting example is the elementary cellular automaton with rule 22, which behaves as a fractal in Wolfram's class 2 for a certain segment of initial configurations, followed by phase transitions of more disordered evolutions typical of class 3 for certain other initial configurations. One question therefore is whether it is class 2 or class 3 behaviour that dominates rule 22 for a greater length of time and a larger number of initial configurations. An analytical answer provides a clear clue: only highly symmetric initial configurations produce fractal-like behaviour (Class 2), but the longer the string the smaller the fraction of highly symmetric strings, therefore the more unlikely the fractal behaviour. Empirically, as we will quantify using the $NC$ measure, the fraction of non-compressed evolutions of rule 22 increases non-linearly (apparently $O(\log)$) as the number of initial configurations increase with $NC$ slowly converging to 1.

There is significant agreement between the classifications produced by Shannon entropy, block compression ($NC$) and algorithmic probability ($BDM$), these last two being approximations of Kolmogorov complexity. Table~\ref{classification} depicts the classification order for each of the measures.

\begin{table}\centering
\ra{1.1}
  \tbl{The 18 most complex ECAs according to the BDM (including the 16 Class 3 and 4 ECAs according to Wolfram's classification) sorted by BDM and Block entropy and BDM normalised by lossless compression ratio. The agreement among the 3 is very high. The Pearson correlation coefficient between Block entropy and lossless compression $NC$ values is 0.936, 0.80 between $NC$ and BDM, and 0.86 between BDM and Block entropy (see Fig~\ref{correlationgraph}).}
{\begin{tabular}{@{}rrrr@{}}\\[-2pt]
\botrule
\multicolumn{1}{c}{ECA} & \multicolumn{1}{c}{Block}  &  \multicolumn{1}{c}{compression} &  \multicolumn{1}{c}{BDM}\\
Rule no. & entropy & ratio ($NC$) &  \multicolumn{1}{c}{value}  \\ \hline
30 & 0.02774 & 1.00 & 0.681\\ 
45 & 0.02705 & 1.00 & 0.642\\ 
54 & 0.02462 & 0.799 & 0.2225\\ 
60 & 0.02430 & 0.783 & 0.1273\\ 
73 & 0.02398 & 1.00 & 0.3515\\ 
90 & 0.02363 & 0.786 & 0.1920\\ 
110 & 0.02329 & 1.00 & 0.3813\\ 
150 & 0.02320 & 0.978 & 0.2585\\ 
94 & 0.02285 & 0.742 & 0.2416\\ 
105 & 0.02270 & 1.00 & 0.2728\\ 
146 & 0.02234 & 0.841 & 0.2990\\ 
126 & 0.02223 & 0.844 & 0.3200\\ 
18 & 0.02199 & 0.836 & 0.2757\\ 
22 & 0.02177 & 0.878 & 0.3279\\ 
122 & 0.02135 & 0.809 & 0.2424\\ 
62 & 0.01919 & 0.969 & 0.2709\\ 
106 & 0.01766 & 0.599 & 0.2801\\
41 & 0.01744 & 0.910 & 0.1832\\
\botrule
\end{tabular}}
  \label{classification}
\end{table}

\begin{figure}[h!]
\begin{center}
\scalebox{.5}{\includegraphics{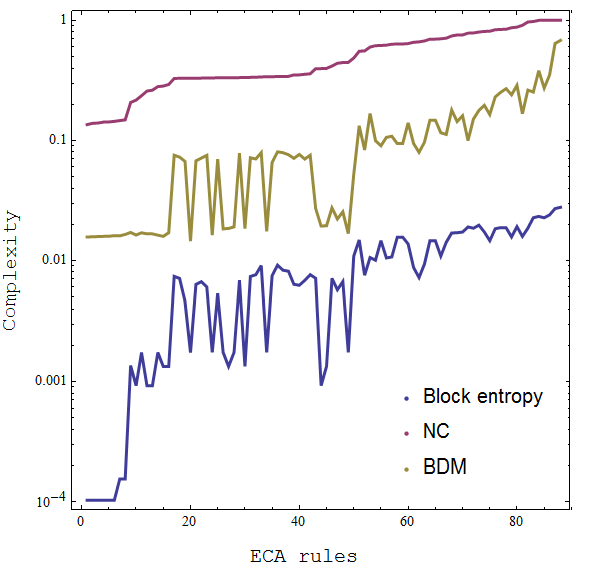}}
\end{center}
\caption{Agreement between the complexity values for the 88 non-equivalent ECA  according to Block entropy, $NC$ and BDM. Rules sorted from smallest to largest $NC$ value. Full ECA tables are given in Figs.~\ref{ECAsEntropy}, \ref{ECAsBDM} and~\ref{ECAsCompressibility} in the Appendix.}
\label{correlationgraph}
\end{figure}

\subsection{Block Decomposition Method}

Another method for approximating the Kolmogorov complexity of $d$-dimensional objects was advanced in~\cite{kolmo2d}. The method is called Block Decomposition (BDM) and is based on the Coding Theorem Method~\cite{d4,d5} (CTM) rooted in the relation established by algorithmic probability between the frequency of production of a string and its Kolmogorov complexity. The more frequent the less random, according to the Coding theorem:

\begin{equation}
\label{coding}
D(s) = \Sigma_{p:U(p) = s} 1/2^{|p|}
\end{equation}
That is, the sum over all the programs for which $U$ with $p$ outputs $s$ and halts on a (prefix-free) universal Turing machine $U$.

Then we define the Kolmogorov complexity $K_m$ of an object $s$ as follows:
\begin{equation}
K_m(s) = -\log_2(\mathbb{D}(s))
\end{equation}

Where $\mathbb{D}(s)$ is the frequency of $s$ as calculated in~\cite{kolmo2d}. The BDM is then defined by:

\begin{equation}
\label{newecaeq}
K^\prime_m(s(i,t)) = \sum_{(r_u,n_u)\in s(i,t)_{d\times d}} \log_2(n_u) + K_m(r_u)
\end{equation}

\noindent Where $r_u$ are the different square arrays in the partition matrix $s(i,t)_{d\times d}$ and $n$ the number of $r_u$ square arrays $d\times d$ from the space-time evolution of a CA $s(i,t)_{d\times d}$. 

A normalised version of the BDM will be used, defined as follows:

\begin{equation}
\label{nbdm}
K^\prime_m(s(i,t))/|s(i,t)|
\end{equation}

Where $|s(i,t)|$ is the size (length $\times$ width) of the space-time diagram produced by the CA $s$ for initial configuration $i$ and runtime $t$, this latter being the length. The reader should simply bear in mind from now on that $K^\prime_m$ (from now on also only denoted by $K_m$) is the approximation of Kolmogorov complexity by the Block Decomposition Method (BDM).

The BDM is used as a complement and an alternative to lossless compression, given that compression algorithms do not deal effectively with small objects. More on algorithmic probability and the Coding theorem can be found in~\cite{cover,calude}, while more on applying the Coding theorem to numerically approximate Kolmogorov complexity and BDM can be found in~\cite{d4,d5,numerical,kolmo}.

\section{Estimation of fractions of Wolfram classes in CA rule spaces}

The behavioural classification of all ECAs according to the current Wolfram classification, as given in~\cite{wolfram}, was possible after only 8 initial configurations. That is, all ECAs reached their assumed Wolfram class $W(s)$ using the measure $NC(s)$ after exploring only 8 initial configurations according to the Gray-code enumeration scheme.

The universality of Kolmogorov complexity as a measure of complexity guarantees that if a cellular automaton has regularities these regularities will eventually be detected. This is why the formalisation of Wolfram's classes in a mathematical function--given in terms of a limit proposed in Section~\ref{wolframclassification}--provides lower bounds on their complexity and is well defined.

\begin{figure}[h!]
\begin{center}
\scalebox{.55}{\includegraphics{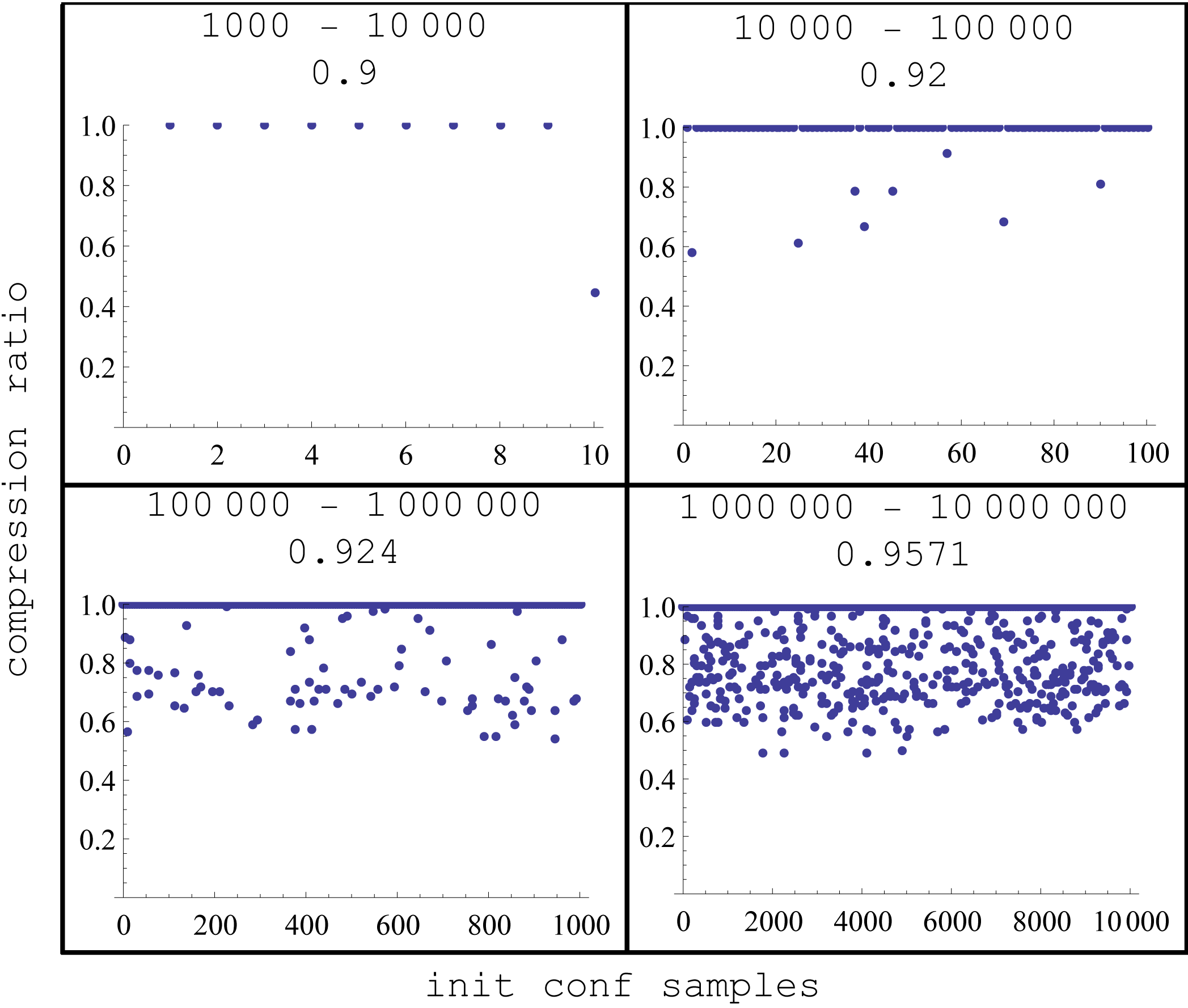}}
\end{center}
\caption{Asymptotic behaviour of Rule 22: Monotonic increase of number and percentage of uncompressed evolutions when the order of magnitude is exponentially increased. Each bin contains $(10^{n+2}-10^{n + 1})/1000$ rule evolutions starting from different initial configurations in the interval  $10^{n+1}$ to $10^{n+2}$ according to the Gray-code enumeration. Fractions of complex behaviour versus simple behaviour for different initial configurations range from 0.9 to 0.9571.}
\label{rule22asymp}
\end{figure}

\begin{figure}[h!]
\begin{center}
\scalebox{.42}{\includegraphics{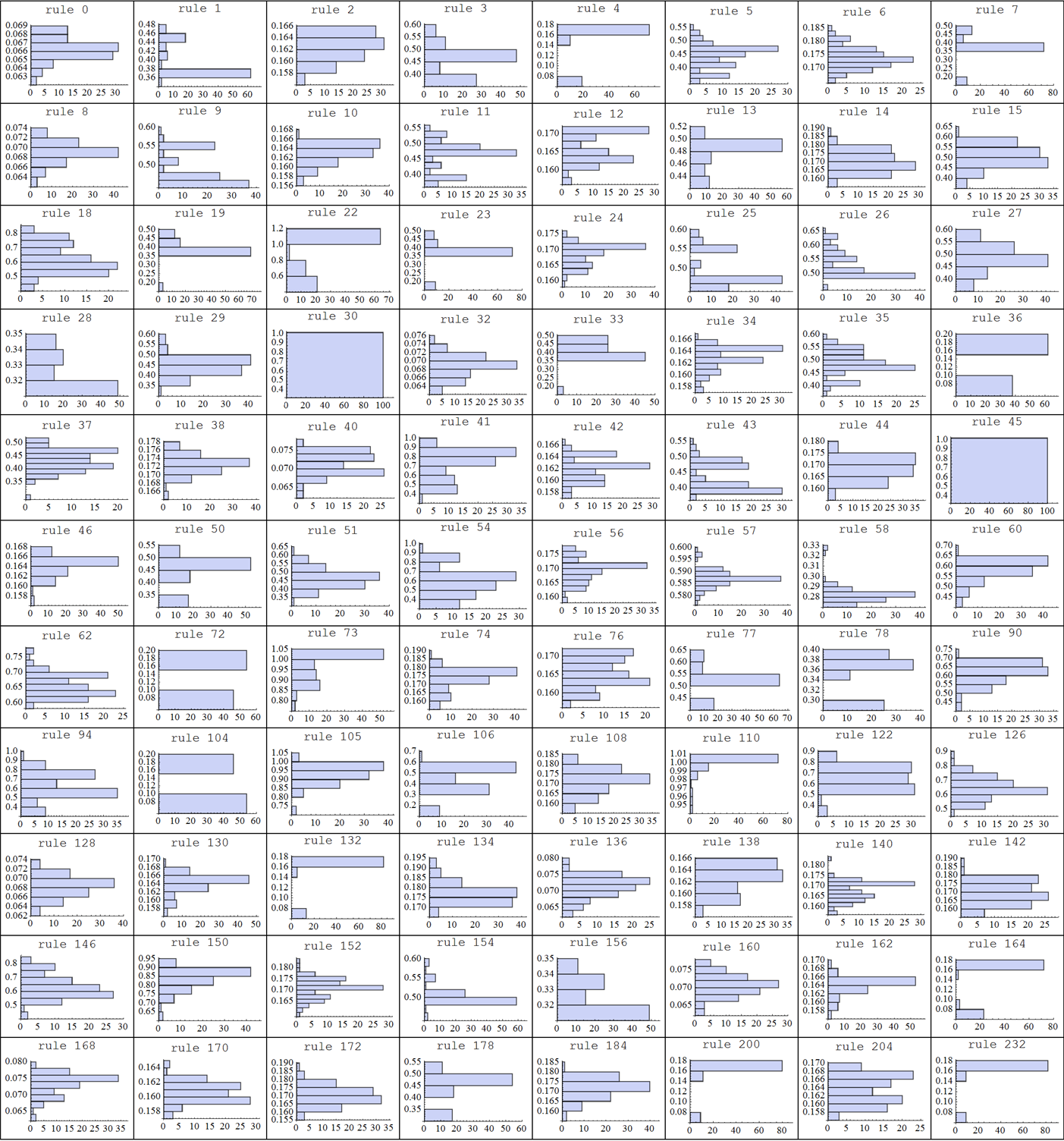}}
\end{center}
\caption{Histograms of uncompressible evolutions (according to $NC$) of all the 88 non-equivalent ECAs for 1000 initial configurations following the Gray-code enumeration scheme, each running for $t=100$ steps. In keeping with what is known about ECA, examining the distribution shapes in these histograms allows one to ascertain which ones are complex and to which Wolfram class they belong. For example, rules known to be in the same Wolfram class exhibit similar compressibility distributions.}
\label{allecas}
\end{figure}

The distributions in Fig.~\ref{allecas} show many interesting details of the (un)compressible evolutions of each of the 88 non-equivalent ECAs under left-right reflection, exchanging states 0 and 1, or by performing these two transformations sequentially. For example, rules in Wolfram's Class 4 are negatively skewed, indicating an asymptotic behaviour towards incompressibility. Class 2, however, produces more Gaussian-like evolutions with high mean compressibility, while almost all the evolutions of rules in Class 1 are maximally compressible. Rules known for their random (or chaotic) behaviour in Class 3 show maximal incompressibility, which is consonant both with what is known and with their Wolfram class.

Rules 30 and 45 show maximum incompressibility and therefore have a high estimated Kolmogorov complexity. Rule 110 displays the maximum asymptotic incompressibility and rules 41, 60, 62, 73 and 94 show a high degree of incompressibility. $NC$ characterises three other ECA rules (rules 62, 73 and 94) whose histograms of their uncompressed evolutions suggest they are not obviously in classes 1 or 2 (hence potentially in either classes 3 or 4), therefore the rest 70 ECA rules would belong to classes 1 or 2. Tables~\ref{classestable} and ~\ref{classestable2} summarise these results.

\begin{table}\centering
\ra{1.1}
  \tbl{Statistical properties from the histograms of the 18 ECA rules in (see Fig.~\ref{allecas}) that bring all 15 ECA rules in classes 3 and 4 according to Wolfram. Rules 30 and 45 have maximal complexity and therefore indeterminate ($i$) skewness. The skewness, distribution mean and standard deviation parameters that characterise all ECA rules in Wolfram's classes 3 and 4 according to Wolfram~\citeyear{wolfram} is given by mean greater than 0.59 or, skewness smaller than -0.87 and standard deviation greater than 0.55, or standard deviation 0 or greater than 0.1.}
{\begin{tabular}{@{}rrrrc@{}}\\[-2pt]
\botrule
\multicolumn{1}{c}{ECA} & \multicolumn{1}{c}{skewness}  &  \multicolumn{1}{c}{mean} &  \multicolumn{1}{c}{standard} &  \multicolumn{1}{c}{Wolfram}\\
\multicolumn{1}{c}{Rule no.} &   &   &  \multicolumn{1}{c}{deviation} &  \multicolumn{1}{c}{Class}\\ \hline
18 & 0.319 & 0.619 & 0.099 & 3\\
22 & -0.78 & 0.854 & 0.203 & 3\\
30 & $i$ & 1 & 0 & 3\\
41 & -0.515 & 0.723 & 0.146 & 4\\
45 & $i$ & 1 & 0 & 3\\
54 & 0.329 & 0.598 & 0.149 & 4\\
60 & -1.12 & 0.58 & 0.0507 & 3\\
62 & 0.602 & 0.657 & 0.037 & 2\\
73 & -1.164 & 0.955 & 0.0596 & 2\\
90 & -0.728 & 0.614 & 0.0597 & 3\\
94 & -0.085 & 0.621 & 0.145 & 2\\
105 & -1.089 & 0.927 & 0.0550 & 3\\
106 & -0.875 & 0.44 & 0.121 & 2\\
110 & -3.1 & 0.995 & 0.0116 & 4\\
122 & -0.544 & 0.65 & 0.106 & 3\\
126 & 0.148 & 0.64 & 0.0772 & 3\\
146 & 0.409 & 0.63 & 0.0848 & 3\\
150 & -1.004 & 0.834 & 0.0589 & 3\\
\botrule
\end{tabular}}
  \label{classestable}
\end{table}

\begin{table}\centering
\ra{1.1}
  \tbl{Suggested behavioural classification of the 88 non-symmetric ECA rules according to the statistical properties of the $NC$ distributions (see Fig.~\ref{allecas}) for each class and the logical/statistical relation given in Table~\ref{classestable} that retrieves all Wolfram's class 3 and 4 ECAs according to Wolfram and which brings 3 other ECA rules that according to the $NC$ histograms are not obviously in classes 1 or 2.}
{\begin{tabular}{@{}cl@{}}\\[-2pt]
\botrule
\multicolumn{1}{c}{Wolfram} &  \\
\multicolumn{1}{c}{Class} &  Non-symmetric ECA rules\\ \hline
$C_{3,4}$: & 18, 22, 30, 41, 45, 54, 60, 62, 73, 90, 94, 105, 106, 110, 122, 126\\
& 146, 150\\ \hline
$C_{1,2}$: & 0, 1, 2, 3, 4, 5, 6, 7, 8, 9, 10, 11, 12, 13, 14, 15, 19, 23, 24, 25, 26, \\
				&  27, 28, 29, 32, 33, 34, 35, 36, 37, 38, 40, 42, 43, 44, 46, 50, 51, 56, \\
				& 57, 58, 72, 74, 76, 77, 78, 104, 108, 128, 130, 132, 134, 136, 138, \\
				& 140, 142, 152, 154, 156, 160, 162, 164, 168, 170, 172, 178, 184,\\
				& 200, 204, 232\\
\botrule
\end{tabular}}
 \label{classestable2}
\end{table}

\begin{table}\centering
\ra{1.1}
  \tbl{Complexity ratios according to Kolmogorov complexity approximated by the normalised Block Decomposition Method (NBDM) based on algorithmic probability (the Coding theorem). Increasing $r$ led to a greater fraction of complex rules in the case of rules with mean values larger than 0.20 NBDM among the first 20 initial configurations according to the Gray-code enumeration, with each rule running for $t=50$ steps. Where we proceeded by sampling, sample sizes are indicated in parentheses.}
{\begin{tabular}{@{}cccc@{}}\\[-2pt]
\botrule
\multicolumn{1}{c}{$CA$} & \multicolumn{1}{c}{NBDM for}  &  \multicolumn{1}{c}{Total number of rules} \\
\multicolumn{1}{c}{range} ($r$)& \multicolumn{1}{c}{$k = 2$}    &  \multicolumn{1}{c}{($2^{2^{2r+1}}$)} \\ \hline
 $1 $ & $0.14$ & 256 (ECA) \\
 $3/2$ &  $0.3241$ & 65\,536\\
 2 &  0.5206 (10\,000) & 4\,294\,967\,296 \\
$5/2$ & 0.795 (10\,000) & 18\,446\,744\,073\,709\,551\,616 \\
\botrule
\end{tabular}}
  \label{bdmtable}
\end{table}

For general CA rules the values shown in Table~\ref{bdmtable} obtained using the normalised Block Decomposition Method (see Eq.~\ref{nbdm}) to approximate Kolmogorov complexity are consistent with the results from Shannon entropy (Table~\ref{entropytable} and Table~\ref{entropytable2}) and compressibility (Table~\ref{maintable}). Space-time diagrams were found to show increasingly larger fractions of higher entropy and algorithmic randomness according to all these complexity measures.

\begin{table}\centering
\ra{1.1}
\tbl{Block Shannon entropy increases for increasing $r$ among totalistic CA ($CA_T$) rules. Block Shannon entropy is averaged over the first 20 initial configurations according to the Gray-code enumeration, each rule running for $t=50$ steps. Only rules with Block Shannon entropy larger than .20 are counted as \emph{complex}.}
{\begin{tabular}{@{}cccc@{}}\\[-2pt]
\botrule
\multicolumn{1}{c}{$CA_T$} & \multicolumn{1}{c}{}  &  \multicolumn{1}{c}{Number of rules} \\
\multicolumn{1}{c}{range ($r$)} & \multicolumn{1}{c}{$k = 2$}  &  \multicolumn{1}{c}{($k^{(k - 1)(2 r + 1) + 1}$)} \\ \hline
 $1$ & 0.125 & 16 \\
 $3/2$ & 0.25 & 32 \\
 2 & 0.39 & 64 \\
$5/2$ & 0.4375 & 128 \\
3 & 0.53125 & 256 \\
$7/2$ & 0.568 & 512 \\
4 & 0.6054 & 1024 \\
$9/2$ & 0.6279 & 2048 \\
5 & 0.6547 & 4096 \\
$11/2$ & 0.676 & 8192 \\
6 & 0.694 & 16\,384 \\
13/2 & 0.708 & 32\,768 \\
\botrule
\end{tabular}}
  \label{entropytable}
\end{table}

\begin{table}\centering
\ra{1.1}
  \tbl{Block Shannon entropy also increased for $k=3$ among totalistic CA rule spaces. Block Shannon entropy is averaged over the first 20 initial configurations according to the Gray-code enumeration, each rule running for $t=50$ steps. Only rules with Block Shannon entropy larger than .20 are counted as \emph{complex}.}
{\begin{tabular}{@{}cccc@{}}\\[-2pt]
\botrule
\multicolumn{1}{c}{$CA_T$} & \multicolumn{1}{c}{}  &  \multicolumn{1}{c}{Number of rules} \\
\multicolumn{1}{c}{range ($r$)} & \multicolumn{1}{c}{$k = 3$}  &  \multicolumn{1}{c}{($k^{(k - 1)(2 r + 1) + 1}$)} \\ \hline
 $r = 1$ & 0.329 & 2187 \\
 $3/2$ & 0.504 & 19\,683 \\
 2 & 0.605 & 177\,147 \\
\botrule
\end{tabular}}
  \label{entropytable2}
\end{table}

\begin{table}\centering
\ra{1.1}
  \tbl{Ratios of compressibility/uncompressibility in totalistic $CA_T$ rule spaces of neighbourhood $r$ and number of symbols $k$. Increasing both $r$ and $k$ increases the fraction of uncompressible evolutions among the first 20 initial configurations according to the Gray-code enumeration, with each running for $t=50$ steps. Where we proceeded by sampling, sample sizes are indicated in parentheses.}
{\begin{tabular}{@{}cccc@{}}\\[-2pt]
\botrule
\multicolumn{1}{c}{$CA_T$ range ($r$)} & \multicolumn{1}{c}{$k = 2$}  &  \multicolumn{1}{c}{ $k = 3$} \\ \hline
 $1$ & $6/16 = 0.375$ & $1059/2187 = 0.484$ \\
 $3/2$ & $15/32 = 0.468$ & $ 672/1000 = 0.672\textit{ }(19\,683)$\\
 2 & 28/64 = 0.4375 & 658/886 = 0.742 (177\,147) \\
$5/2$ & 63/128 = 0.492 & $6487/7972 = 0.813$ (1\,594\,323) \\
3 & 146/256 = 0.57 & \\
$7/2$ & $313/512 = 0.611$ & \\
4 & $665/1024 = 0.649$ & \\
$9/2$ & $1398/2048 = 0.682$ & \\
5 & $2911/4096 = 0.7$ & \\
$11/2$ & $6041/8192 = 0.737$ & \\
6 & $12\,461/16\,384 = 0.76$ & \\
13/2 & $25\,562/32\,768 = 0.78$ & \\
\botrule
\end{tabular}}
  \label{maintable}
\end{table}

\begin{figure}[h!]
\begin{center}
\scalebox{.3}{\includegraphics{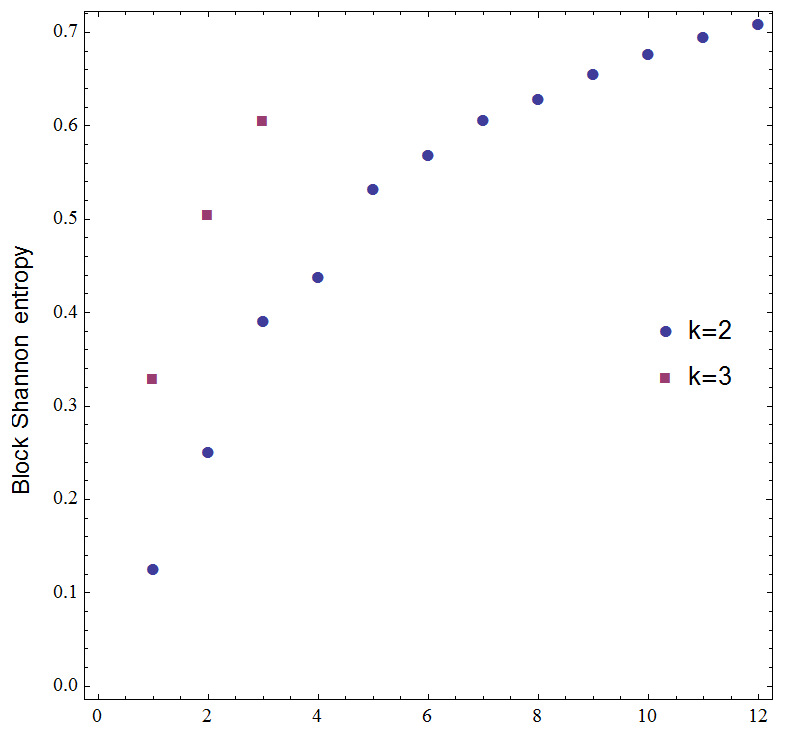}}\scalebox{.3}{\includegraphics{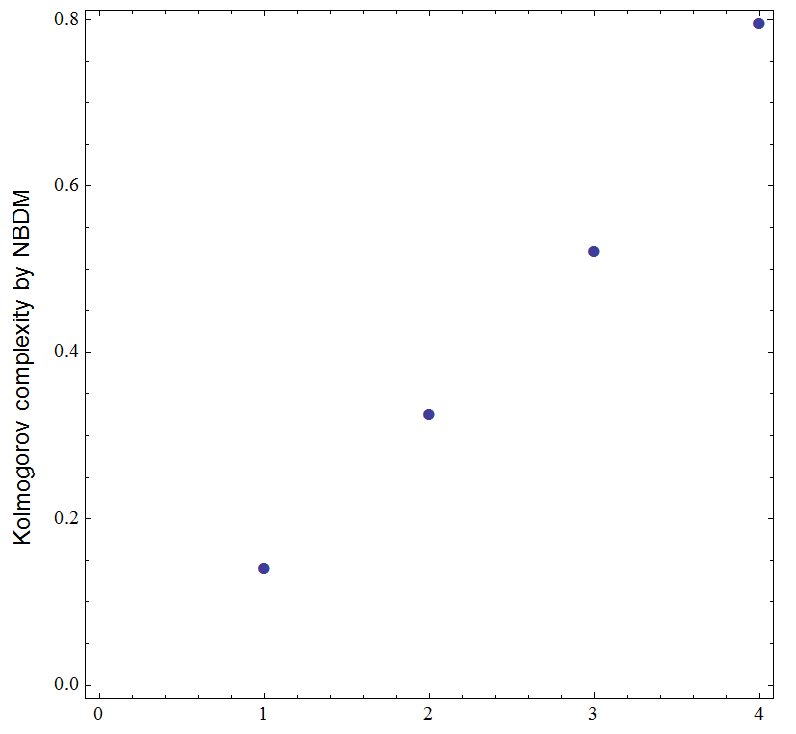}}
\end{center}
\caption{Left: Block Shannon entropy progression in different rule spaces for increasing $r$ and $k$. Right: Kolmogorov complexity progression for $k=2$ according to the NBDM based on algorithmic probability. For $k=3$, $r=5/2$ already has a larger fraction of complex systems than $k=2,r=13/2$.}
\label{mainfig2}
\end{figure}

\begin{figure}[h!]
\begin{center}
\scalebox{.33}{\includegraphics{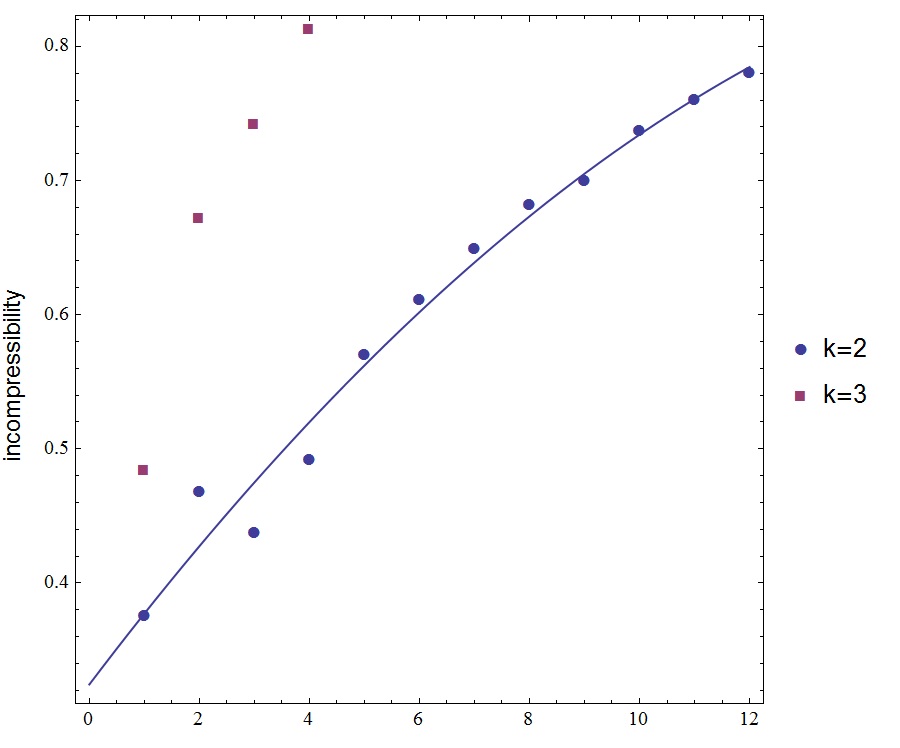}}
\end{center}
\caption{Plot of compressibility ratios ($NC$) in different rule spaces for increasing $k$ and $r$ in totalistic CA rules spaces for $k = 2$ and 3. The curve that fits the data points for $k = 2$ is a logarithmic line.}
\label{mainfig}
\end{figure}

Table~\ref{maintable} shows the results of the exhaustive search in which a rule was considered incompressible (and therefore \emph{complex}) if at least one initial configuration led to a space-time evolution asymptotically incompressible among the first 50 steps. The rule space for increasing $k$ grows too fast, so we proceeded by sampling for an initial segment of $k = 3$. Fig.~\ref{mainfig} and Fig.~\ref{mainfig2} show the main results of the complexity experiments. For each evolution the compression ratio ($NC$) was calculated. Those CA with an evolution having a compression ratio greater than 1/2 were considered incompressible for at least one initial configuration up to the number of time steps chosen. Below that threshold CAs were considered simple.

The rate of convergence to 1 of $CA_T$ seems to follow a quadratic equation. Given that rule spaces are proper supersets of smaller rule spaces, the number of complex rules cannot be 1, but Fig.~\ref{mainfig} shows the growth rate of the function and the speed of its apparent asymptotic movement towards 1. 

When sets were not exhausted, as reported in Table~\ref{maintable}, sample sizes were calculated by running several samples of increasing size. A sample of 0.005 the size of the original rule space gave us a complex/simple ratio with 2 precision digits. Increasing the sample size by 10, hence by 0.05 (or 5\%) of the total rule space size, increased the accuracy by only an extra decimal digit. It was therefore decided that a size of 0.005 provided enough statistical confidence.

\section{Characterisation of Wolfram's Classification in terms of Algorithmic Randomness }

The present study offers a new and complementary formulation of Wolfram's classes, producing a number of interesting properties to look into. The following is a characterisation of Wolfram's classes in terms of asymptotic behaviour and (un)compressibility based on algorithmic information theory:

\begin{itemlist}
\item {Class 1.} Evolution has asymptotic compressibility ratio equal to 0 (lowest Kolmogorov complexity).
\item {Class 2.} Evolution has compressibility ratio smaller than 1/2 (low Kolmogorov complexity).
\item {Class 3.} Evolution has compressibility ratio 1. (uncompressible randomness)
\item {Class 4.} Evolution has asymptotic compressibility ratio equal to 1. (asymptotic uncompressible randomness)
\end{itemlist}

This is useful in determining the difficulty of deciding the Wolfram class of a dynamic system, implying that Wolfram's classification is itself uncomputable in this formal version, a direct inference from the uncomputability of $K$.

\section{Conclusions}

We have studied the asymptotic behaviour of cellular automata both for initial configurations and runtimes, as well as the fractions of Wolfram classes for increasingly larger rule spaces of computer programs, specifically cellular automata. Among the results, we confirmed preliminary findings reporting~\cite{wolfram2d} a preponderance of complex classes 3 and 4 as the number of symbols (colours) and neighbourhood ranges of these systems increase. An open question is whether the same trend persists for larger dimensions (or other computing models) although these authors have no evidence to believe that the trend will be different.

We have shown that behaviour is for the most part stable for a given initial condition but that complexity increases in terms of the number of both the initial configurations and rules spaces. The rate at which it does so is fast before asymptotically slowing down approaching compressibility ratio 1 (uncompressibility), growing by $\log(r^(1/k))$ after regression analysis, with $r$ the neighbourhood and $k$ the number of colours. Similar results were obtained by using two different measures. One was Shannon's entropy in the form of a block entropy version normalised and applied to CA space-time diagrams. The other measure was Kolmogorov complexity approximated by 2 different alternative methods, the traditional lossless compression algorithm and the Coding theorem based on algorithmic probability, from which a Block decomposition technique was devised.

While the Shannon entropy result may be less surprising because the number of possible configurations with more symbols and larger neighbourhood ranges suggests an increase in entropy, the fact that the 3 approaches used to measure complexity in rule spaces (including estimations of a universal and objective measure such as Kolmogorov complexity) yielded the same results provides a robust answer to the question regarding the change of fractions of behaviour and the rate of this change in computer programs for increasing resources where complexity is clearly seen to dominate asymptotically.

\newpage

\appendix{}

\begin{figure}[h!]
\begin{center}
\scalebox{.6}{\includegraphics{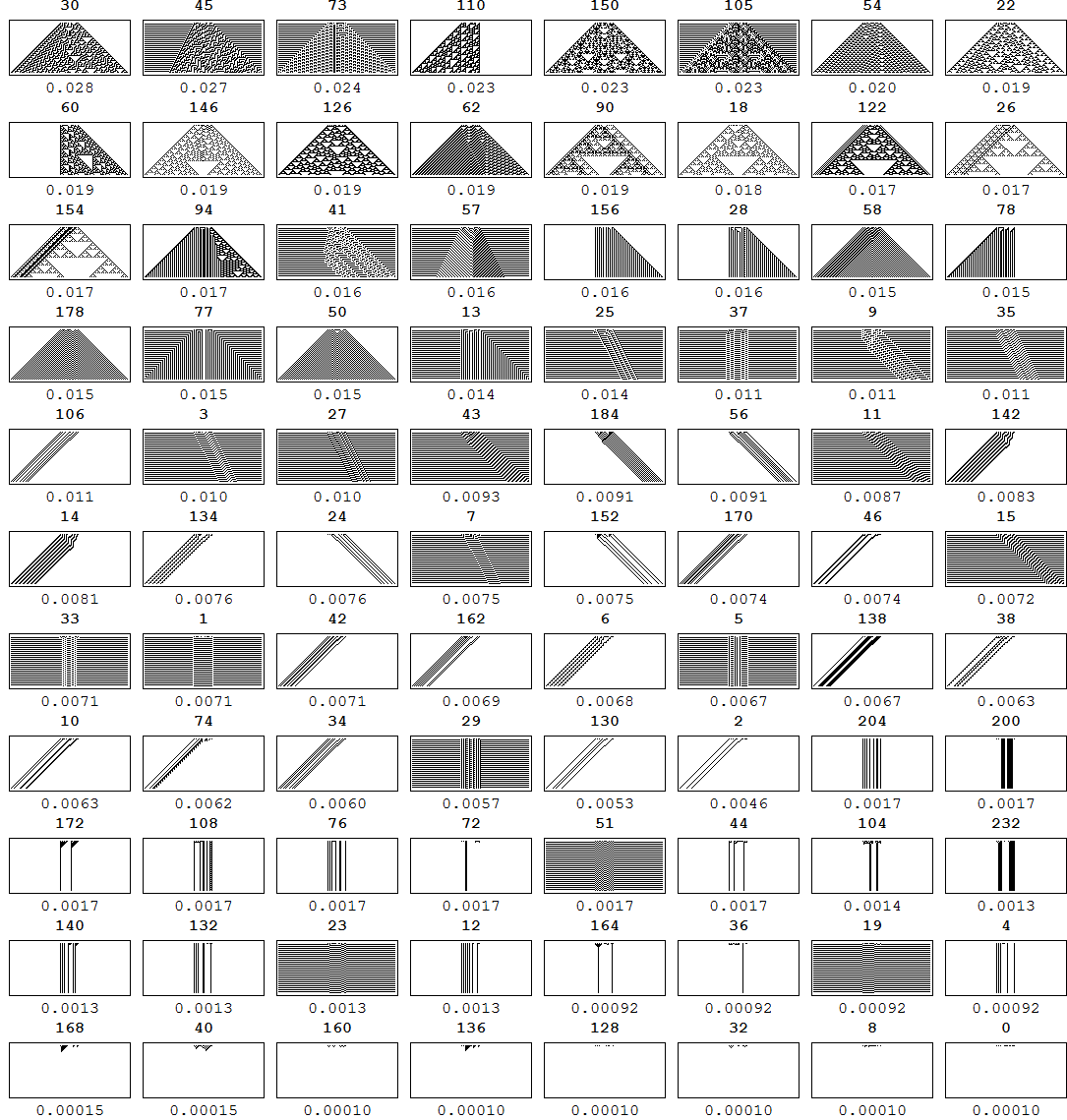}}
\end{center}
\caption{The 88 different ECA sorted from largest to smallest mean Block entropy over the first 20 initial configurations according to the Gray-code enumeration, normalised by $t \times N^N$ with $N$ as the chosen Entropy block size (in this case 3 and 4) and $t=50$ the runtime. Each ECA is shown here starting from a random initial configuration.}
\label{ECAsEntropy}
\end{figure}

\newpage

\begin{figure}[h!]
\begin{center}
\scalebox{.6}{\includegraphics{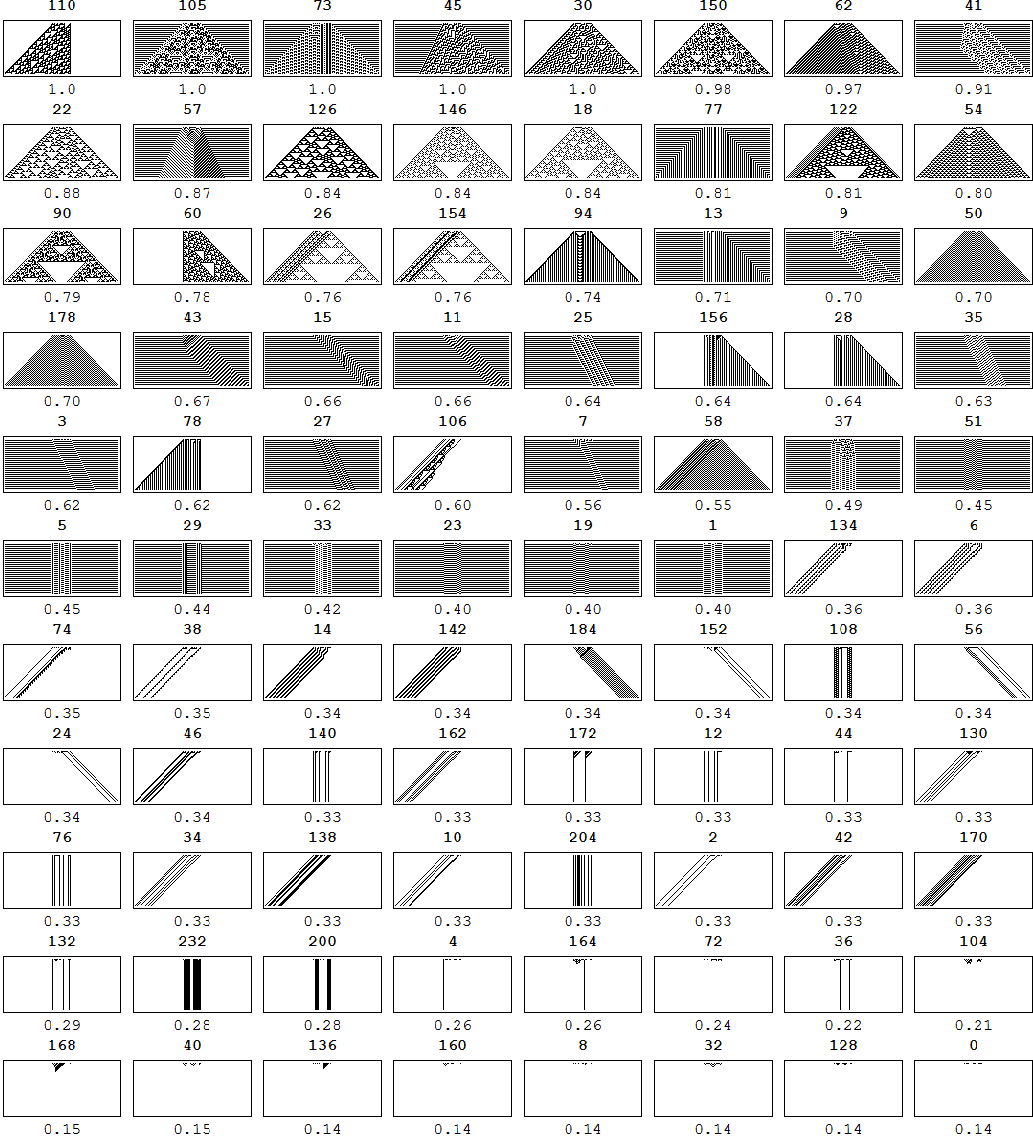}}
\end{center}
\caption{The 88 different ECA sorted from largest to smallest mean lossless compression over the first 20 initial configurations according to the Gray-code enumeration. For example, rules 110 (124), 105, 73 (109), 45 (75, 89, 101) and 30 (86) are among the less compressible (hence the most algorithmically complex). Each ECA is shown here starting from a random initial configuration.}
\label{ECAsCompressibility}
\end{figure}

\newpage

\begin{figure}[h!]
\begin{center}
\scalebox{.56}{\includegraphics{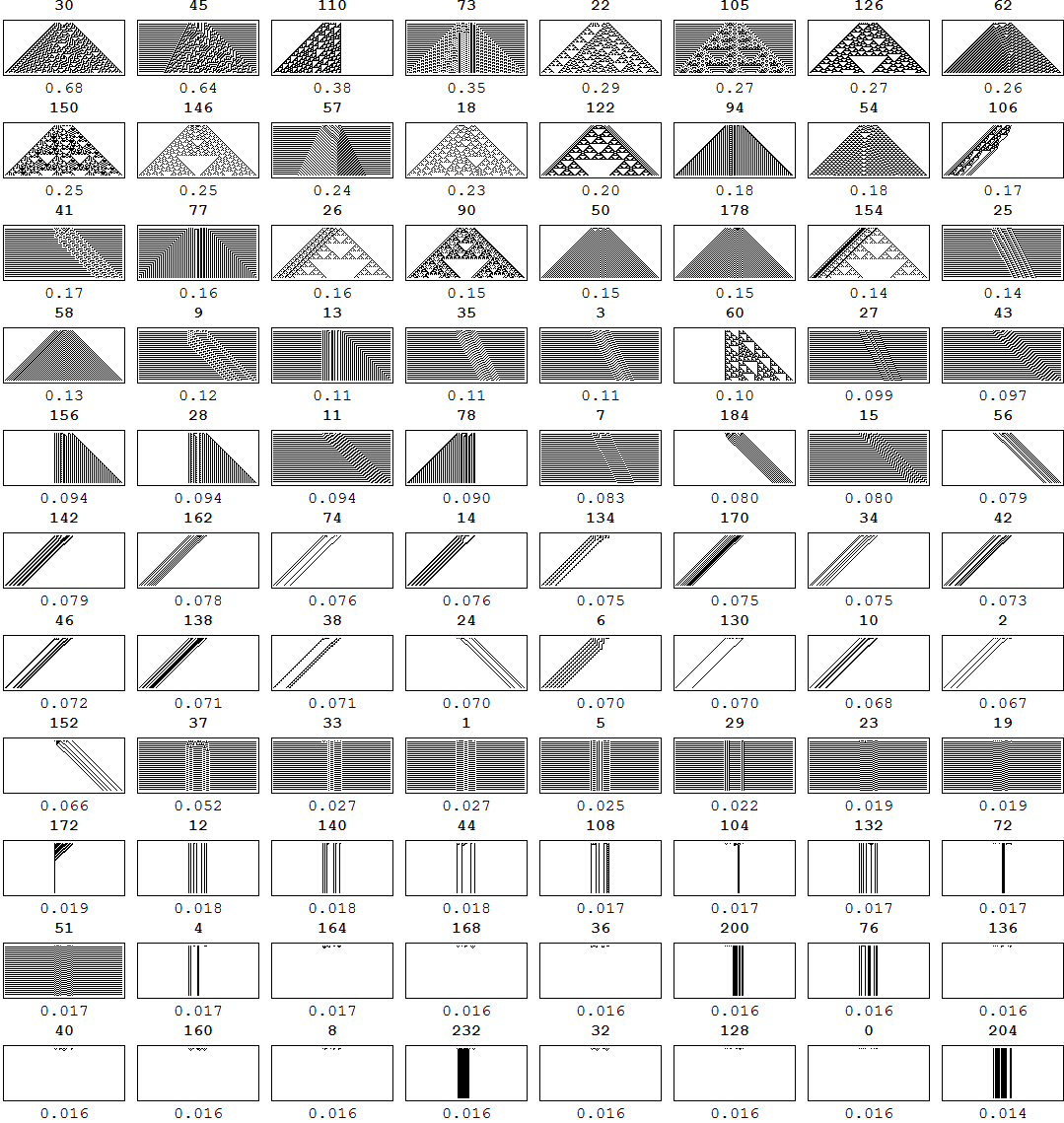}}
\end{center}
\caption{The 88 different ECA sorted from largest to smallest normalised mean Block Decomposition (BDM) over the first 20 initial configurations according to the Gray-code enumeration.}
\label{ECAsBDM}
\end{figure}


\newpage

\begin{thebibliography}{9}

\bibitem[Calude(2002)]{calude} C.S. Calude. [2002] {\it Information and Randomness,} (Springer, Heidelberg, Germany).
\bibitem[Chaitin(1980)]{chaitin} G.J. Chaitin. [1980] {\it Information, Randomness and Incompleteness: Papers on Algorithmic Information Theory,} (World Scientific, Singapore).
\bibitem[Cook(2004)]{cook} M. Cook. [2004] ``Universality in Elementary Cellular Automata", {\it Complex Systems,} 15, pp.~1--40.
\bibitem[Cover \& Thomas(2006)]{cover} T.M. Cover \& J.A. Thomas. [2006] {\it Information Theory,} (J. Wiley and Sons, USA).
\bibitem[Culik II \& Yu(1988)]{kn:CY88} K. Culik II, and S. Yu. [1988] ``Undecidability of CA Classification Schemes", {\it Complex Systems,} 2(2) 177--190.
\bibitem[Delahaye \& Zenil(2012)]{d4} J.-P. Delahaye \& H. Zenil. [2012] ``Numerical Evaluation of the Complexity of Short Strings: A Glance Into the Innermost Structure of Algorithmic Randomness", {\it Applied Math. and Comp}.
\bibitem[G\'acs(1974)]{gacs} P. G\'acs. [1974] ``On the symmetry of algorithmic information", {\it Soviet Mathematics Doklady,} 15:1477--1480.
\bibitem[Gardner(1970)]{life} M. Gardner. [1970] ``Mathematical Games - The fantastic combinations of John Conway's new solitaire game `life'", pp. 120--123, {\it Scientific American} 223.
\bibitem[Gr\"unwald \& Vit\'anyi(2004)]{grunwald} P. Gr\"unwald \& P. Vit\'anyi. [2004] ``Shannon Information and Kolmogorov Complexity", arXiv:cs/0410002 [cs.IT].
\bibitem[Gutowitz(1991)]{gutowitz} H. Gutowitz (Ed.) [1991] {\it Cellular Automata: Theory and Experiment,} (MIT Press, USA).
\bibitem[Kolmogorov(1965)]{kolmo} A.N. Kolmogorov. [1965] ``Three approaches to the quantitative definition of information", {\it Problems of Information and Transmission}, 1(1):1--7.
\bibitem[Langton(1991)]{langton} C.G. Langton. [1991] ``Life at the Edge of Chaos". In C.G. Langton et al. {\it Artificial Life II,} (Addison-Wesley, USA).
\bibitem[Martinez {\it et al.}(2013)]{genaronew} G. J. Martinez, A. Adamatzky, R. Alonso-Sanz, ``Designing Complex Dynamics with Memory", \emph{In preparation (personal communication)}.
\bibitem[Shannon(1948)]{shannon} C. Shannon. [1948] ``A mathematical theory of communication," {\it Bell System Technical Journal,} 27:379-423, 623--656.
\bibitem[Schnorr(1971)]{schnorr} C.P. Schnorr. [1971] ``A unified approach to the definition of a random sequence," {\it Mathematical Systems Theory} 5 (3): 246--258.
\bibitem[Li \& Vit\'anyi(2008)]{li} M. Li \& P. Vit\'anyi. [2008] \emph{An Introduction to Kolmogorov Complexity and Its Applications,} (Springer, Heidelberg, Germany).
\bibitem[Martin-L\"of(1966)]{martin} P. Martin-L\"of. [1966] ``The Definition of Random Sequences", {\it Information and Control,} 9(6): 602--619.
\bibitem[Martinez {\it et al.}(2013)]{genaroca} G.J. Martinez, J.C. Seck-Touh-Mora \& H. Zenil. [2012] ``Computation and Universality: Class IV versus Class III Cellular Automata," {\it J. of Cellular Automata},  vol. 7, no. 5-6, p. 393--430.
\bibitem[von Neuman(1966)]{neumann} J. von Neumann; A.W. Burks (ed.). [1966] {\it Theory of self-reproducing automata,} (Urbana, University of Illinois Press, USA).
\bibitem[Packard \& Wolfram(1985)]{wolfram2d} N.H. Packard \& S. Wolfram. [1985] ``Two-Dimensional Cellular Automata", {\it Journal of Statistical Physics,} vol. 38:5-6, pp. 901--946.
\bibitem[Soler-Toscano {\it et al.}(2012)]{d5} F. Soler-Toscano, H. Zenil, J.-P. Delahaye and N. Gauvrit. [2012] ``Calculating Kolmogorov Complexity from the Frequency Output Distributions of Small Turing Machines", arXiv:1211.1302 [cs.IT].
\bibitem[Soler-Toscano et al.(2012b)]{numerical} F. Soler-Toscano, H. Zenil, J.-P. Delahaye and N. Gauvrit. [2012] ``Correspondence and Independence of Numerical Evaluations of Algorithmic Information Measures", arXiv:1211.4891 [cs.IT].
\bibitem[Wuensche(1997)]{wuensche} A. Wuensche. [1997] ``Attractor Basins of Discrete Networks; Implications on self-organisation and memory", {\it Cognitive Science Research Paper 461,} Univ. of Sussex, D.Phil thesis.
\bibitem[Wuensche(2005)]{wuensche2} A. Wuensche. [2005] ``Glider dynamics in 3-value hexagonal cellular automata: the beehive rule", {\it Int. Journ. of Unconventional Computing,} vol.1, no.4, 375--398.
\bibitem[Wolfram(1985)]{wolfram1985} S. Wolfram. [1985] ``Twenty Problems in the Theory of Cellular Automata", {\it Physica Scripta}, T9 170--183.
\bibitem[Wolfram(2002)]{wolfram} S. Wolfram [2002] {\it A New Kind of Science}, (Wolfram Media, Champaign, IL. USA).
\bibitem[Wolfram(2002b)]{wolframopen} S. Wolfram. [2002] ``A New Kind of Science: Open Problems and Projects", \url{http://www.wolframscience.com/openproblems/NKSOpenProblems.pdf},  (accessed on December 2012).
\bibitem[Zenil(2010)]{zenilca} H. Zenil. [2010] ``Compression-based Investigation of the Dynamical Properties of Cellular Automata and Other Systems", {\it Complex Systems.} 19(1), pages 1--28.
\bibitem[Zenil(2012)]{kn:ZenAISB}  H. Zenil. [2012] ``Nature-like Computation and a Measure of Computability". In G. Dodig-Crnkovic R. \& Giovagnoli (eds), {\it Natural Computing/Unconventional Computing and its Philosophical Significance}, SAPERE Series, (Springer, Heidelberg, Germany).
\bibitem[Zenil(2012b)]{kn:Zen12} H. Zenil. [2011] ``On the Dynamic Qualitative behaviour of Universal Computation", {\it Complex Systems,} vol. 20, No. 3, 265--278.
\bibitem[Zenil(2012c)]{acomputableuniverse} H. Zenil (ed.). [2012] {\it A Computable Universe}, (World Scientific Publishing Press, Singapore).
\bibitem[Zenil et al.(2012d)]{kolmo2d} H. Zenil, F. Soler-Toscano, J.-P. Delahaye \& N. Gauvrit. [2012] ``Two-Dimensional Kolmogorov Complexity and Validation of the Coding Theorem Method by Compressibility", arXiv:1212.6745 [cs.CC].
\bibitem[Zvonkin and Levin(1970)]{levin} A. K. Zvonkin \& L. A. Levin. [1970] ``The complexity of finite objects and the development of the concepts of information and randomness by means of the theory of algorithms", {\it Russian Mathematical Surveys,} 25(6):83--124.

\end{thebibliography}
\end{document}